\documentclass{article}
\usepackage{spconf,amsmath,graphicx}
\usepackage{graphicx}
\usepackage{amsmath}
\usepackage{amsfonts}
\usepackage{amssymb}
\usepackage[amssymb]{SIunits}
\usepackage{longtable}%flexible table definitions

\usepackage{import}%to allow import of svg drawings
\usepackage{color}%to import svg drawings in color
\usepackage{amsthm}
\usepackage{cite}
\usepackage{subcaption}
\usepackage{textcomp}
\usepackage{algorithm}
\usepackage{algorithmic}
\usepackage{lipsum}% http://ctan.org/pkg/lipsum
\usepackage{multicol}% http://ctan.org/pkg/multicols
\usepackage{multirow}

\title{Asynchrounous Decentralized Learning of a Neural Network}

\name{Xinyue Liang, Alireza M. Javid, Mikael Skoglund, Saikat Chatterjee}
	
\address{
	Division of Information Science and Engineering \\                    
	School of Electrical Engineering and Computer Science, KTH Royal Institute of Technology, Sweden                 \\
	xinyuel@kth.se, almj@kth.se, skoglund@kth.se, sach@kth.se
	}
\begin{document}
\ninept

	
	%Telephone: (800) 555--1212, Fax: (888) 555--1212}


	\maketitle
	
\begin{abstract}
		In this work, we exploit an asynchronous computing framework namely ARock to learn a deep neural network called self-size estimating feedforward neural network (SSFN) in a decentralized scenario. Using this algorithm namely asynchronous decentralized SSFN (dSSFN), we provide the centralized equivalent solution under certain technical assumptions. Asynchronous dSSFN relaxes the communication bottleneck by allowing one node activation and one side communication, which reduces the communication overhead significantly, consequently increasing the learning speed. We compare asynchronous dSSFN with traditional synchronous dSSFN in the experimental results, which shows the competitive performance of asynchronous dSSFN, especially when the communication network is sparse.
\end{abstract}
	
	

	% \fi
	%    \IEEEpeerreviewmaketitle

	\vspace{-5pt}
	\section{Introduction}
	Decentralized learning has received a high interest in signal processing, machine learning and data analysis \cite{Kamp2018,Decentralized_neural_network_2,Decentralized_signal_processing,Decentralized_learning_1,Decentralized_learning_2,Decentralized_learning_3,Decentralized_learning_4}. Privacy, security and unavailability of all data in a single place are primary reasons for decentralized learning where data are distributed over several places (or nodes). In a decentralized scenario, nodes are not allowed to share data and there is no master node that has access to all nodes. 
	
	An ideal objective of a decentralized learning algorithm is to provide a centralized equivalence: parameters learned in the decentralized setup should converge to the corresponding centralized setup. Besides, a decentralized algorithm should have low complexity (computational complexity and communication overhead) and work for asynchronous information exchange. It has practical advantages - processing in a node does not need to wait for all neighbor nodes to pass information, nodes can have varying processing powers, etc. 
	
	The design and learning of a neural network is a non-convex optimization problem. Gradient descent and its versions, such as stochastic gradient descent are typically applied for parameter optimization of the neural network. Gradient descent for a neural network can be realized in a decentralized setup where training data is distributed over nodes. A recent study on convergence of decentralized gradient descent for non-convex optimization problems found that it is non-trivial to establish the centralized equivalence (see proposition 1 and discussion afterward in \cite{Nonconvex_gradient_descent}). The work \cite{ADMM_NN2016}  employs another method to train the neural networks using alternating-direction-method-of-multipliers (ADMM) which provides an approximate solution to the non-convex problem, therefore, it cannot gaurantee centralized equivalence.
	
	Our main contribution in this article is to design a multi-layer neural network with the following requirements: it can learn over a decentralized setup using asynchronous information exchange and the learning requires a low complexity. We investigate the scope of a recently proposed multi-layer feed-forward neural network \cite{saikat_SSFN} that is found to be suitable for satisfying all the requirements. The weight matrices of the neural network are the parameters to be learned. The weight matrices are learned using a layer-by-layer learning approach. A weight matrix is a combination of two submatrices: one submatrix is a random matrix instance and the other submatrix is an outcome of a convex optimization problem. Due to the use of convex optimization, we can employ ADMM \cite{ADMM_Bertsekas, Boyd_ADMM_2011}. ADMM is a well-known algorithm for the decentralized solution of a convex optimization problem with the centralized equivalence. ADMM can also be realized for asynchronous information exchange. An appropriate combination of decentralized ADMM and layer-by-layer learning approach leads to the design of a neural network with the centralized equivalence.
	
	\noindent \textbf{Relevant literature review.} The term `decentralized' often comes with the term `distributed'. There exists a significant amount of research on developing distributed neural networks \cite{Dean_NIPS2012,distributed_neural_network_1,distributed_neural_network_2,distributed_neural_network_3,distributed_neural_network_4,distributed_neural_network_5,distributed_neural_network_6}. They use a master-slave architecture that helps to provide the centralized equivalence. For instance, The work \cite{Dean_NIPS2012} proposes distributed neural networks for model parallelism and data parallelism scenarios where a master node coordinates information exchange in asynchronous manner. 
	
	There exist works using decentralized optimization methods for neural networks where there is no master node \cite{Compressed_GD2018,Lian2017,ADMM_NN2016,DPLN2018}. The work \cite{Compressed_GD2018} uses amplified-differential compression for computing the gradient in a synchronous manner. It has the advantage of low computational complexity. In the work \cite{Lian2017}, pair-wise information exchange is required where a node has to transmit and receive information from at least one neighbor node in an iteration of the decentralized learning process.
	
	ARock is an asynchronous computing algorithm that can handle decentralized convex optimization problems \cite{ARock}. ARock does not require pair-wise information exchange. It allows even a relaxed condition - one side (single) activation and one side communication. That means one node can update with out-of-date information and send updates to neighbors without prior notice. ARock has low complexity. We explore the use of ARock for ADMM implementation in our development of the decentralized neural network.  
	
	\vspace{-5pt}
	\section{Decentralized SSFN}
	
	We begin this section with a decentralized problem formulation for a feedforward neural network. Then we briefly explain the architecture and learning of a  (centralized) specific neural network that is amenable for our requirements. Finally, we provide decentralized and asynchronous training. 
	\vspace{-7pt}
	\subsection{Decentralized problem}
	\vspace{-5pt}
	In a supervised learning problem, let $(\mathbf{x}, \mathbf{t})$ be a pair-wise form of input data vector $\mathbf{x}$ that we observe and target vector $\mathbf{t}$ that we wish to infer. We assume $\mathbf{x}\in\mathbb{R}^{P}$ and $\mathbf{t}\in\mathbb{R}^{Q}$. A neural network is an inference function providing output $\tilde{\mathbf{t}} = \mathbf{f}(\mathbf{x},\boldsymbol{\theta})$, where $\boldsymbol{\theta}$ contains the parameters of the neural network. Let $(\mathbf{x}^{(j)},\mathbf{t}^{(j)})$ denotes the $j$'th data-and-target pair in a $J$ sample training dataset $\mathcal{D}=\{ (\mathbf{x}^{(j)},\mathbf{t}^{(j)}) \}_{j=1}^J$. We can use a cost function
	\begin{equation}
	C(\boldsymbol{\theta}) = \sum_{(\mathbf{x}^{(j)},\mathbf{t}^{(j)}) \in \mathcal{D}} \|\mathbf{t}^{(j)} - \mathbf{f}(\mathbf{x}^{(j)}, \boldsymbol{\theta}) \|_2^2.
	\end{equation}
	For a centralized scenario, we have all the $J$ samples in a single node. In the training phase, we learn optimal parameters by minimizing the cost in a regularized manner, as follows
	\begin{equation}
	\boldsymbol{\theta}_c^{\star} = \underset{\boldsymbol{\theta} }{\mathrm{argmin}} \, C(\boldsymbol{\theta}) \,\, \mathrm{s.t.\,\,} \| \boldsymbol{\theta} \|_2^2 \leq \epsilon,
	\label{eq:CentralizedProblem}
	\end{equation}
	where `s.t.' is the abbreviation of `subject to'.
	The above problem is non-convex for a general neural network and hence, non-trivial. 
	
	In a distributed scenario, we have $M$ processing nodes and we assume that the training dataset $\mathcal{D}$ is divided into $M$ datasets. We denote the training dataset at the $m$'th node by $\mathcal{D}_m$ such that $\cup_m \mathcal{D}_m = \mathcal{D}$. Let us denote the cost at the $m$'th node by
	\begin{equation}
	C(\boldsymbol{\theta}_m) = \sum_{(\mathbf{x}^{(j)},\mathbf{t}^{(j)}) \in \mathcal{D}_m} \|\mathbf{t}^{(j)} - \mathbf{f}(\mathbf{x}^{(j)}, \boldsymbol{\theta}_m) \|_2^2
	\end{equation}
	where $\boldsymbol{\theta}_m$ denotes the parameters learned at the $m$'th node. Our interest is to learn the parameters in a distributed manner as follows
	\begin{equation}
	\boldsymbol{\theta}_d^{\star} \! = \!  \underset{\boldsymbol{\theta},\boldsymbol{\theta}_{m}}{\mathrm{argmin}} \, \sum_m \! C(\boldsymbol{\theta}_m)  \,\, \mathrm{s.t. \,\,} \forall m, \, \boldsymbol{\theta}_m \! = \! \boldsymbol{\theta},\, \| \boldsymbol{\theta} \|_2^2 \leq \epsilon.
	\label{eq:DeCentralizedProblem}
	\end{equation}
	The constraint $\forall m, \boldsymbol{\theta}_m \! = \! \boldsymbol{\theta}$ enforces the same solution for all nodes. In the next subsection we discuss a neural network for which we we can achieve $\boldsymbol{\theta}_c^{\star} = \boldsymbol{\theta}_d^{\star}$ using ADMM under some technical conditions. This is the centralized equivalence.

	\vspace{-7pt}
	\subsection{A neural network}
	\vspace{-5pt}
	The neural network we investigate in this article is called self-size estimating feedforward neural network (SSFN) \cite{saikat_SSFN}. The main motivation for developing SSFN in \cite{saikat_SSFN} is the self-estimation of its size, which means the network automatically finds the necessary number of layers and number of nodes. The SSFN training algorithm is shown to have a low computational requirement due to its layer-by-layer learning approach. The SSFN does not use gradient search (backpropagation) and hence avoids the requirement of a high computational resource. It is also shown in \cite{saikat_SSFN} that the optimization of weight matrices using backpropagation does not lead to significant performance improvement. In our work in this article, we refrain from estimating the size of the decentralized setup due to a high search complexity. Instead, we design a fixed-size decentralized feedforward neural network based on the SSFN learning principle such that we retain the complexity advantage. We assume that each hidden layer has a fixed number of hidden neurons. Even if we are not estimating its size, it is doable at the expense of high complexity.
	
\begin{figure*}[t!]
		\centering
		\def\svgwidth{\linewidth}
		\includegraphics[width=1\textwidth]{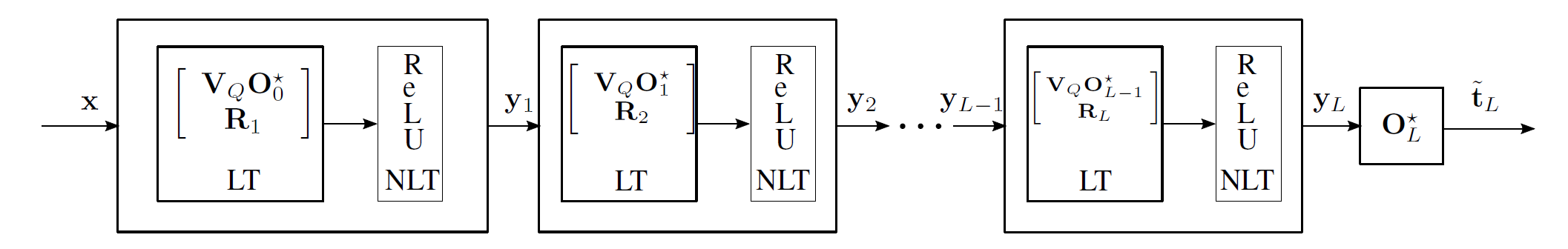}
		\vspace{-5pt}
		\caption{The architecture of a multi-layer SSFN with $L$ layers and its signal flow diagram. LT stands for \emph{linear transform} (weight matrix) and NLT stands for \emph{non-linear transform} (activation function). We use ReLU activation function.}
		\label{fig:MultiLayerPLN}
		\vspace{-17pt}
	\end{figure*}

	The SSFN signal flow relation is
	\begin{eqnarray*}
		\tilde{\mathbf{t}} = \mathbf{O} \mathbf{g}(\mathbf{W}_L \, \mathbf{g}(\hdots \mathbf{g}(\mathbf{W}_2 \, \mathbf{g}(\mathbf{W}_1 \, \mathbf{x}))\hdots)) = \mathbf{Oy},
	\end{eqnarray*}
	where $\mathbf{W}_l$ are weight matrices and $\mathbf{g}(.)$ denotes the non-linear transform comprised of rectified-linear-unit (ReLU) activation function. The parameters to learn are $\pmb{\theta} = \{ \{ \mathbf{W}_l \}, \mathbf{O} \}$. Figure \ref{fig:MultiLayerPLN} shows the architecture of SSFN with its weight matrices. 
	\medmuskip=-2mu
	\medmuskip=4mu
	
	%While the work of {\color{red}{\cite{}}} developed the SSFN architecture that learns its parameters and size of the network automatically, we work with a fixed size SSFN and learn its parameters. We refrain from estimating its size as it requires a high complexity. The number of layers $L$ and the hidden neurons for the $l$'th layer $n_l$ are fixed a priori. We also assume that all layers have the same number of hidden neurons, which means $n_l =n, \forall l$.
	
	The SSFN addresses the optimization problem \eqref{eq:CentralizedProblem} in a suboptimal manner. The SSFN parameters $\mathbf{O}$ and $\{\mathbf{W}_l\}$  are learned layer-by-layer as a sequential forward learning approach. The feature vector of $l$'th layer is 
	\begin{eqnarray}
	\mathbf{y}_l = \mathbf{g}(\mathbf{W}_l \, \mathbf{g}(\hdots \mathbf{g}(\mathbf{W}_2 \, \mathbf{g}(\mathbf{W}_1 \, \mathbf{x}))\hdots)) \in \mathbb{R}^n. 
	\end{eqnarray}
	The layer-by-layer sequential learning approach starts with layer number $l=0$ and then the layers are increased one-by-one until we reach $l=L$. We have $\mathbf{y}_0 = \mathbf{x}$. Let us first assume that we have an $l$-layer network. The $(l+1)$-layer network will be built on the $l$-layer network. For designing the $(l+1)$-layer network given the $l$-layer network, the steps for finding parameter $\mathbf{W}_{l+1}$ at the $(l+1)$'th layer are as follows:
	\begin{enumerate}
		\item For all the samples in the training dataset $\mathcal{D}$, we compute $\mathbf{y}_l^{(j)} = \mathbf{g}(\mathbf{W}_l \, \mathbf{g}(\hdots \mathbf{g}(\mathbf{W}_2 \, \mathbf{g}(\mathbf{W}_1 \, \mathbf{x}^{(j)}))\hdots))$.
		\item Using the samples $\{ \mathbf{y}_l^{(j)} \}_{j=1}^J$ we solve the following convex optimization problem 
		%\begin{eqnarray}
		%\mathcal{C}_l = \sum_{j=1}^J  \| \mathbf{t}^{(j)} - \mathbf{O}_l \, \mathbf{y}_l^{(j)} \|^2.
		%\end{eqnarray} 
		%We compute the optimal regression matrix $\mathbf{O}_l$ by solving the convex optimization problem
		\begin{eqnarray}
		\begin{array}{l}
		\mathbf{O}_{l}^{\star} \!\! = \! \underset{ \mathbf{O}_{l}  }{\arg\min}  \! \displaystyle \sum_{j=1}^J  \! \|  \mathbf{t}^{(j)} \! - \! \mathbf{O}_l \, \mathbf{y}_l^{(j)} \! \|^2  \,\,  \mathrm{s.t.} \,
		\| \mathbf{O}_l \|_F^2 \! \leq \! \epsilon_l.
		\end{array}
		\label{eq:LayerWise_ConvexOptimization}
		\end{eqnarray}
		It is shown in \cite{saikat_SSFN} that we can choose the regularization parameters $\epsilon_l=\epsilon=2Q, \, l=0,1,2,\hdots,L.$ 
		%{\color{blue}{The choice for $\epsilon_{0}$ requires cross-validation}}. 
		Note that $\mathbf{O}_0$ is $Q \times P$-dimensional, and every $\mathbf{O}_l$ for $l=1,2,\hdots,L$ is $Q \times n$-dimensional. 
		\item We form the weight matrix for the $(l+1)$'th layer
		\begin{equation}
		\mathbf{W}_{l+1}=
		\left[
		\begin{array}{c}
		\mathbf{V}_Q \mathbf{O}_{l}^{\star} \\ \mathbf{R}_{l+1}
		\end{array}
		\right],
		%\nonumber
		\label{eq:weightMatrix}
		\end{equation}
		where $\mathbf{V}_Q=[\mathbf{I}_Q \,\, -\mathbf{I}_Q]^{T}$ is a fixed matrix of dimension $2Q \times Q$, $\mathbf{O}_{l}^{\star}$ is learned by convex optimization \eqref{eq:LayerWise_ConvexOptimization}, and 
		$\mathbf{R}_{l+1}$ is an instance of random matrix. The matrix $\mathbf{R}_0$ is $(n-2Q) \times P$-dimensional, and every $\mathbf{R}_l$ for $l=1,2,\hdots,L$ is $(n-2Q) \times n$-dimensional. Note that we only learn $\mathbf{O}_{l}^{\star}$ to form $\mathbf{W}_l$. We do not learn $\mathbf{R}_{l}$ and it can be pre-fixed before training of SSFN. Deciding the weight matrix, the $(l+1)$-layer network is
		$
		\mathbf{y}_{l+1} = \mathbf{g}(\mathbf{W}_l \, \mathbf{g}(\hdots \mathbf{g}(\mathbf{W}_2 \, \mathbf{g}(\mathbf{W}_1 \, \mathbf{x}))\hdots)) \\
		= \mathbf{g}(\mathbf{W}_l \mathbf{y}_{l})
		$.
	\end{enumerate}
	It is shown in \cite{saikat_SSFN} that the three steps mentioned above guarantees monotonically decreasing cost $\sum_{j}  \| \mathbf{t}^{(j)} - \mathbf{O}_l \, \mathbf{y}_l^{(j)} \|^2$ with increasing the layer number $l$. The monotonically decreasing cost with increase in layer number is the key to address the optimization problem \eqref{eq:LayerWise_ConvexOptimization} as we continue to add new layers one-by-one and set the weight matrix of a layer using \eqref{eq:weightMatrix}. It was experimentally shown (see Table 5 of \cite{saikat_SSFN}) that the use of gradient search (backpropagation) for further optimization of parameters in SSFN could not provide a reasonable performance improvement. Note that backpropagation based optimization requires a significant computational complexity.
	
	\vspace{-7pt}
	\subsection{Decentralized SSFN for Asynchronous Communication}
	\vspace{-5pt}
	\label{subsec:Decentralized_SSFN_using_ADMM_Asynchronous}

	%    \medmuskip=-2mu
	%    \begin{figure*}[ht!]
	%        \centering
	%        \def\svgwidth{\linewidth}
	%        \input{Fig_I_3.eps_tex}    
	%        \caption{The architecture of a multi-layer SSFN with $L$ layers. `LT' stands for linear transform and `NLT' stands for non-linear transform.}
	%        \vspace{-10pt}
	%        \label{fig:MultiLayerPLN}
	%    \end{figure*}
	%    \medmuskip=4mu
	
	Let us consider that the decentralized setup of $M$ nodes is expressed by a set of undirected connections between the nodes. The connection set is $\mathit{E}=\{ (m,n) | \mathrm{if}\,\,\mathrm{node}\,\,m\,\,\mathrm{connects\,\,to\,\,node}\,\,n\}$. The $m$'th node has $J_m$ training samples, that means $| \mathcal{D}_m | = J_m$. Let us introduce the observation data matrix $\mathbf{X}_m \in \mathbb{R}^{P \times J_m}$ and the target data matrix $\mathbf{T}_m \in \mathbb{R}^{Q \times J_m}$ using the dataset $\mathcal{D}_m$. Here $\mathbf{X}_m$ is formed by the $\mathbf{x}^{(j)} \in \mathcal{D}_m$ and $\mathbf{T}_m$ is formed by the corresponding $\mathbf{t}^{(j)} \in \mathcal{D}_m$. 
	
	Addressing the decentralized problem \eqref{eq:DeCentralizedProblem} that has centralized equivalence to \eqref{eq:CentralizedProblem}, the SSFN design boils down to efficient decentralized solution of \eqref{eq:LayerWise_ConvexOptimization} as follows
	\begin{equation}
	\vspace{-5pt}
	\label{eq:5}
	\begin{aligned}
	& \underset{\mathbf{O}_{l,m}}{\min} \sum_{m=1}^{M} \|\mathbf{T}_m - \mathbf{O}_{l,m}\mathbf{Y}_{l,m} \|_{F}^{2}, \\
	s.t. & \forall (m,n) \in \mathit{E}, \mathbf{O}_{l,m} = \mathbf{O}_{l,n}, \forall m, \|\mathbf{O}_{l,m}\|_F^2 \leq \epsilon, 
	\end{aligned}
	\end{equation}
	where $\mathbf{Y}_{l,m}$ is the signal matrix at the $l$'th layer of SSFN. Note that $\mathbf{Y}_{l,m}$ is generated by feeding the observation data matrix $\mathbf{X}_{m}$ to the $l$'th layer of SSFN. To solve this optimization problem, auxiliary variable $\mathbf{\Lambda}_{mn}$ is introduced associated with each connected link $(m,n)$. Then the original problem can be rewritten as
	\vspace{-5pt}
	\begin{equation*}
	\begin{aligned}
	\underset{\mathbf{O}_{l,m}}{\min} \sum_{m=1}^{M} &\|\mathbf{T}_m - \mathbf{O}_{l,m}\mathbf{Y}_{l,m} \|_{F}^{2}, \,\,\, s.t. \,\,\, \\
	 \forall (m,n) \in \mathit{E}, &  \mathbf{O}_{l,m} = \mathbf{\Lambda}_{mn}, \mathbf{O}_{l,n}=\mathbf{\Lambda}_{mn}, \forall m, \|\mathbf{O}_{l,m}\|_F^2 \leq \epsilon.
	\end{aligned}
	\end{equation*}
	For every pair of constraints $\mathbf{O}_{l,m} = \mathbf{\Lambda}_{mn}$ and $\mathbf{O}_{l,n} = \mathbf{\Lambda}_{mn}$, $\forall (m,n)\in \mathit{E}$, dual variables $\mathbf{Z}_{m,n}$ and $\mathbf{Z}_{n,m}$ are associated to denote the information sent from the $m$'th node to $n$'th node, and vice versa. Applying ARock framework for ADMM iterations, ADMM get simplified as
	\begin{eqnarray*}
	\hspace{-5pt}
	\begin{array}{lll}
%	\mathbf{\Lambda}_{mn}^{(k+1)} = \underset{\mathbf{\Lambda}}{\arg\min} -\hspace{-12pt} \underset{n \in \mathit{N}(m)}{\sum}\hspace{-8pt} \mathbf{Z}_{n,m}^{(k)}\mathbf{\Lambda}  + \gamma\|\mathbf{\Lambda}\|_F^2, \\
	\mathbf{\Lambda}_{mn}^{(k+1)} = -\frac{\mathbf{Z}^{(k)}_{m,n}+\mathbf{Z}^{(k)}_{n,m}}{2\gamma},  \\
	\mathbf{O}_{l,m}^{(k+1)} =\underset{\mathbf{O}}{\arg\min}~ \|\mathbf{T}_m-\mathbf{O}\mathbf{Y}_m\|^2_F+ \hspace{-12pt}\underset{n \in \mathit{N}(m)}{\sum} \hspace{-8pt}\mathbf{Z}_{n,m}^{(k)}\mathbf{O}  \\ \,\,\,\,\,\,\,\,\,\,\,\,\,\,\,\,\,\,\,\,\,\,\, +\frac{\gamma}{2}| \mathit{E(m)}| \cdot \|\mathbf{O}\|_F^2, \,\,s.t.\,\, \|\mathbf{O}\|_F^2 \leq \epsilon,\\
	\mathbf{Z}^{(k+1)}_{m,n} =\mathbf{Z}^{(k)}_{m,n}-\eta(\frac{(\mathbf{Z}^{(k)}_{m,n}+\mathbf{Z}^{(k)}_{n,m})}{2}+\gamma \mathbf{O}_{l,m}^{(k+1)}),  \forall n \in \mathit{N}(m),
	\end{array}
	\end{eqnarray*}
	where $\gamma \in (0,\frac{2}{\|\mathbf{Y}_{l,m}\|})$ is the scalar and $\eta$ is the step size, both of them needed to be chosen properly. $\mathit{E}(m)$ is the set of links connected with node $m$, and $\mathit{N}(m)$ denotes the set of neighbor nodes of the $m$'th node. Note that $\mathbf{\Lambda}_{mn}$ vanishes, $m$'th processing node only updates $\mathbf{O}_{l,m}$ and $\mathbf{Z}_{m,n}$. Therefore when a processing node $m$ is activated, the ADMM iterations are
	\begin{equation}
	\label{eq:7}
	\hspace{-5pt}
	\begin{array}{l}
	\mathbf{O}_{l,m}^{(k+1)}\! =\! \mathcal{P}_{\epsilon}((2\mathbf{T}_{m}\mathbf{Y}_{l,m}^{T} \!\!\!- \hspace{-10pt}\underset{n \in \mathit{N}(m)}{\sum} \hspace{-8pt}\mathbf{Z}_{n,m}^{(k)})(2\mathbf{Y}_{l,m}\mathbf{Y}_{l,m}^{T} \!\!\!+ \!\gamma|\mathit{E(m)}|\mathbf{I})^{-1}),\\
	\mathbf{Z}^{(k+1)}_{m,n}\!=\!\mathbf{Z}^{(k)}_{m,n}-\eta(\frac{(\mathbf{Z}^{(k)}_{m,n}+\mathbf{Z}^{(k)}_{n,m})}{2}+\gamma \mathbf{O}_{l,m}^{(k+1)}),  \forall n \in \mathit{N}(m),\\
	\end{array} 
	\end{equation}
	where $|\mathit{E(m)}|$ is the cardinality of $\mathit{E(m)}$, i.e., the number of connected links associated with node $m$. $\mathbf{Z}_{n,m}^{(k)}$ is the previous received dual variable sent from neighbors $n \in \mathit{N}(m)$ to node $m$. Here $\mathcal{P}_{\epsilon}$ is the projection onto the space of matrices with $\|\cdot\|_F^2\leq\epsilon$ to avoid overfitting, which is
	\begin{equation}
	\mathcal{P}_{\epsilon}(\mathbf{O}_{l,m})=\left\{
	\begin{array}{ll}
	\mathbf{O}_{l,m}\cdot(\frac{\epsilon^{\frac{1}{2}}}{\|\mathbf{O}_{l,m}\|_F}) & : \|\mathbf{O}_{l,m}\|^2_F> \epsilon\\
	\mathbf{O}_{l,m} & : \mbox{otherwise}.
	\end{array}
	\right.
	\end{equation}
	These two steps of ADMM updates would only be executed when some processing node $m$ is activated. This implies that instead of waiting for neighbors to be activated and communicate at the same time, this asynchronous ADMM only requires one side communication during the update. 
	We show the algorithm for this decentralized problem in Algorithm \ref{algorithm}.
%	\vspace{-10pt}

	\begin{algorithm}[ht!]
		\caption{: Algorithm for Asynchronous learning decentralized SSFN (dSSFN)}\label{algorithm}
		\mbox{Input: }
		\begin{algorithmic}[1]
			\STATE Training dataset $\mathcal{D}_m$ for the $m$'th node
			\STATE Parameters to set: $L, \mathrm{number}\,\, \mathrm{of}\,\, \mathrm{hidden}\,\, \mathrm{neurons}, \gamma, \eta$
			\STATE Set of random matrices $\{ \mathbf{R}_l \}_{l=1}^L$ are generated and shared between all nodes 
		\end{algorithmic}
		\mbox{Initialization:}
		\begin{algorithmic}[1]
			\STATE $l=-1$  \hfill (Index for $l$'th layer)
		\end{algorithmic}
		\mbox{Progressive growth of layers:}
		\begin{algorithmic}[1]
			\REPEAT 
			\STATE $l \leftarrow l+1$  \hfill (Increase layer number)
			\STATE $k=0$ \hfill (Iteration index of ADMM)
			\STATE Compute $\mathbf{Y}_{l,m}\! =\! \mathbf{g}(\mathbf{W}_l \hdots \mathbf{g}(\mathbf{W}_1\mathbf{X}_{l,m}))\! =\! \mathbf{g}(\mathbf{W}_l\! \mathbf{Y}_{l-1,m})$ \\
			\hspace{-22pt}\mbox{Solve \eqref{eq:5} using ARock framework to find $\mathbf{O}_{l,m}^{\star}$:}
			%\begin{algorithmic}[1]
			\STATE Each node $m$ initializes the dual variables $\mathbf{Z}_{m,n}^{(0)} = 0$ for $(m,n) \in \mathit{E}(m)$, $K>0$
			\STATE Receive and store $\mathbf{Z}_{n,m}^{(k)}$ from neighbors $n \in \mathit{N}(m)$ at anytime
			\WHILE {$k< K$, any activated processing node $m$}
			\STATE $k \leftarrow k+1$
			\STATE Solve $\mathbf{O}_{l,m}^{(k+1)}$ using \eqref{eq:7}
			\STATE Calculate all the $\mathbf{Z}_{m,n}^{(k+1)} \,\,\,n\in\mathit{N}(m)$ using \eqref{eq:7}
			%; this requires iterative information exchange over graph. Let us assume that the number of iterations for information exchange is $B$. 
			\STATE Send $\mathbf{Z}_{m,n}^{(k+1)}$ to all neighbors
			\ENDWHILE
			%\end{algorithmic}
			\STATE Form weight matrix $\mathbf{W}_{l+1}=
			\left[
			\begin{array}{c}
			\mathbf{V}_Q \mathbf{O}_{l}^{\star} \\ \mathbf{R}_{l+1}
			\end{array}
			\right]$
			\UNTIL  $l = L$
		\end{algorithmic}
		%\mbox{Output:}
		%\begin{algorithmic}
		%\STATE Decentralized 
		%\end{algorithmic}
	\end{algorithm}

	\vspace{-10pt}    
	\section{Experimental Results}
	
	\label{section:SimResult}
	We perform experimental simulations to evaluate the performance of the dSSFN on a circular network structure with undirected connections for classification tasks\footnote{All the Matlab code for experiments are available at https://sites.google.com/site/saikatchatt/.}. The datasets that we use are briefly described in Table \ref{table:Database_for_classification}. The datasets used for the experiments can be found at \cite{DATABASE_1,JIANG_DICT_LEARNING_KSVD_CVPR_2011,Lecun_NORB_online,Lecun_MNIST_online}. We use the $Q$-dimensional target vector $\mathbf{t}$ in a classification task as a discrete variable with an indexed representation of 1-out-of-$Q$-classes. A target variable (vector) instance has only one scalar component that is 1, and the other scalar components are zero. For the local SSFN, we set the maximum number of layers $L=20$, the number of hidden neurons $2Q+1000$, and $\epsilon=2Q$ for each layer. We uniformly divide the training dataset between the processing nodes. We test the algorithm with a different number of processing nodes, and different network degree $d$, i.e., each processing node has $d$ neighbors. The classification performance results are reported in Table \ref{table:Classification_accuracy}, the corresponding parameters we used is shown in Table \ref{table:Classification_parameters}. Our interest is to compare the testing performance of asynchronous dSSFN with traditional synchronous dSSFN and centralized SSFN. 
	
	\begin{table}[t!]
		\centering
		\caption{Databases for multi-class classification}
		\vspace{-10pt}
		\label{table:Database_for_classification}
		\setlength{\tabcolsep}{2.5pt}
		\begin{tabular}{ |c|c|c|c|c| } 
			\hline
			Database & {\begin{tabular}{@{}c@{}}$\#$ of  \\ train data\end{tabular}}  & {\begin{tabular}{@{}c@{}}$\#$ of  \\ test data\end{tabular}} & {\begin{tabular}{@{}c@{}}Input  \\ dimension ($\mathit{P}$)\end{tabular}}  & {\begin{tabular}{@{}c@{}}$\#$ of  \\ classes ($\mathit{Q}$)\end{tabular}} \\
			\hline \hline 
			Vowel & 528 & 462 & 10 & 11 \\ 
			\hline
			Satimage & 4435 & 2000 & 36 & 6 \\ 
			\hline
			Caltech101 & 6000 & 3000 & 3000 & 102 \\ 
			\hline
			Letter & 13333 & 6667 & 16 & 26 \\ 
			\hline
			NORB & 24300 & 24300 & 2048 & 5 \\ 
			\hline
			MNIST & 60000 & 10000 & 784 & 10 \\ 
			\hline
		\end{tabular}
		    \vspace{-7pt}
	\end{table}
	
	\begin{table}[t!]
		\centering
		\caption{Classification performance comparison between centralized SSFN, synchronous dSSFN, and asynchronous dSSFN on a circular graph where $M=20$, $d=8$, $K=200$}
		\vspace{-10pt}
		\label{table:Classification_accuracy}
		\setlength{\tabcolsep}{1.7pt}
		\renewcommand{\arraystretch}{1.1}
		\begin{tabular}{|c|c|c|c|c|c|c|c|c|c|}
			\hline
			\multirow{2}{*}{Dataset} & \multicolumn{1}{c}{central SSFN} & \multicolumn{2}{|c|}{sync dSSFN} & \multicolumn{2}{|c|}{async dSSFN} \\ \cline{2-6}
			&{\begin{tabular}{@{}c@{}}Test \\ Accuracy\end{tabular}} &
			{\begin{tabular}{@{}c@{}}Test \\ Accuracy\end{tabular}} & {\begin{tabular}{@{}c@{}}Training \\ Time(s)\end{tabular}}  & {\begin{tabular}{@{}c@{}}Test \\ Accuracy\end{tabular}} & {\begin{tabular}{@{}c@{}}Training \\ Time(s)\end{tabular}} \\
			\hline \hline 
			
			Vowel & 61.9$\pm$1.7 & 61.3$\pm$1.6 & 34.9 & 61.2$\pm$1.7 & 7.8  \\ 
			\hline
			Satimage & 88.7$\pm$0.5 & 89.2$\pm$0.4 & 16.3 & 89.9$\pm$0.2 & 4.2\\ 
			\hline
			Caltech101 & 74.6$\pm$0.7 & 74.2$\pm$0.6 & 427.5 & 75.4$\pm$0.9 & 27.4 \\ 
			\hline
			Letter & 95.7$\pm$0.2 & 93.1$\pm$0.4 & 43.6 & 95.8$\pm$0.5 & 9.9  \\ 
			\hline
			NORB & 86.2$\pm$1.1 & 83.6$\pm$0.9 & 21.7 & 86.0$\pm$0.5 & 11.9  \\ 
			\hline
			MNIST & 95.8$\pm$0.2 & 94.9$\pm$0.2 & 43.5 & 95.0$\pm$0.3 & 11.7 \\ 
			\hline
		\end{tabular}
		    \vspace{-7pt}
	\end{table}

	\begin{table}[t!]
		\centering
		\caption{The corresponding parameters of Table \ref{table:Classification_accuracy}}
		\vspace{-10pt}
		\label{table:Classification_parameters}
		\setlength{\tabcolsep}{4.8pt}
		\renewcommand{\arraystretch}{1.1}
		\begin{tabular}{ |c|c|c|c|c|c|c| } 
			\hline
			\multirow{2}{*}{Dataset} & \multicolumn{2}{|c|}{central SSFN} &\multicolumn{2}{|c|}{sync dSSFN} & \multicolumn{2}{|c|}{async dSSFN} \\ \cline{2-7}
			& $\mu_0$ & $\mu_l$ & $\mu_0$ & $\mu_l$ & $\gamma_0$  & $\gamma_l$ \\
			\hline \hline
			Vowel & $10^{-5}$ & $10^1$ & $10^{-4}$ & $10^{0}$ & $10^{1.5}$ & $10^{-1}$  \\ 
			\hline
			Satimage & $10^6$ & $ 10^5 $ & $10^{-4}$ & $10^{-1}$ & $10^{0}$ & $10^{-1}$ \\ 
			\hline
			Caltech101 & $10^0$ & $10^{-2}$ & $10^{-1}$ & $10^{0}$ & $10^{0}$ & $10^{-1}$  \\ 
			\hline
			Letter & $10^{2}$ & $10^4$ & $10^{-6}$ & $10^{0}$  & $10^{4}$ & $10^{-2}$  \\ 
			\hline
			NORB & $10^{-8}$ & $10^{2}$ & $10^{-1}$ & $10^{-1}$ & $10^{2}$ & $10^{-2}$ \\ 
			\hline
			MNIST & $10^{2}$ & $10^{5}$ & $10^{-5}$ & $10^0$ & $10^{-1}$ & $10^{-1}$ \\ 
			\hline
		\end{tabular}
		\vspace{-10pt}
	\end{table}

	\noindent\textbf{Centralized equivalent performance.} Based on the numerical results which are provided in Table \ref{table:Classification_accuracy}, one can observe that the synchronous dSSFN and asynchronous dSSFN both provide centralized equivalence solution, while asynchronous dSSFN outperforms traditional synchronous dSSFN in training time. Note that the step size is fixed to be $\eta=0.5$ for every ADMM iteration. 
	
\begin{figure}[t!]
		\centering
%		    \vspace*{-15pt}
		\includegraphics[width=0.4\textwidth]{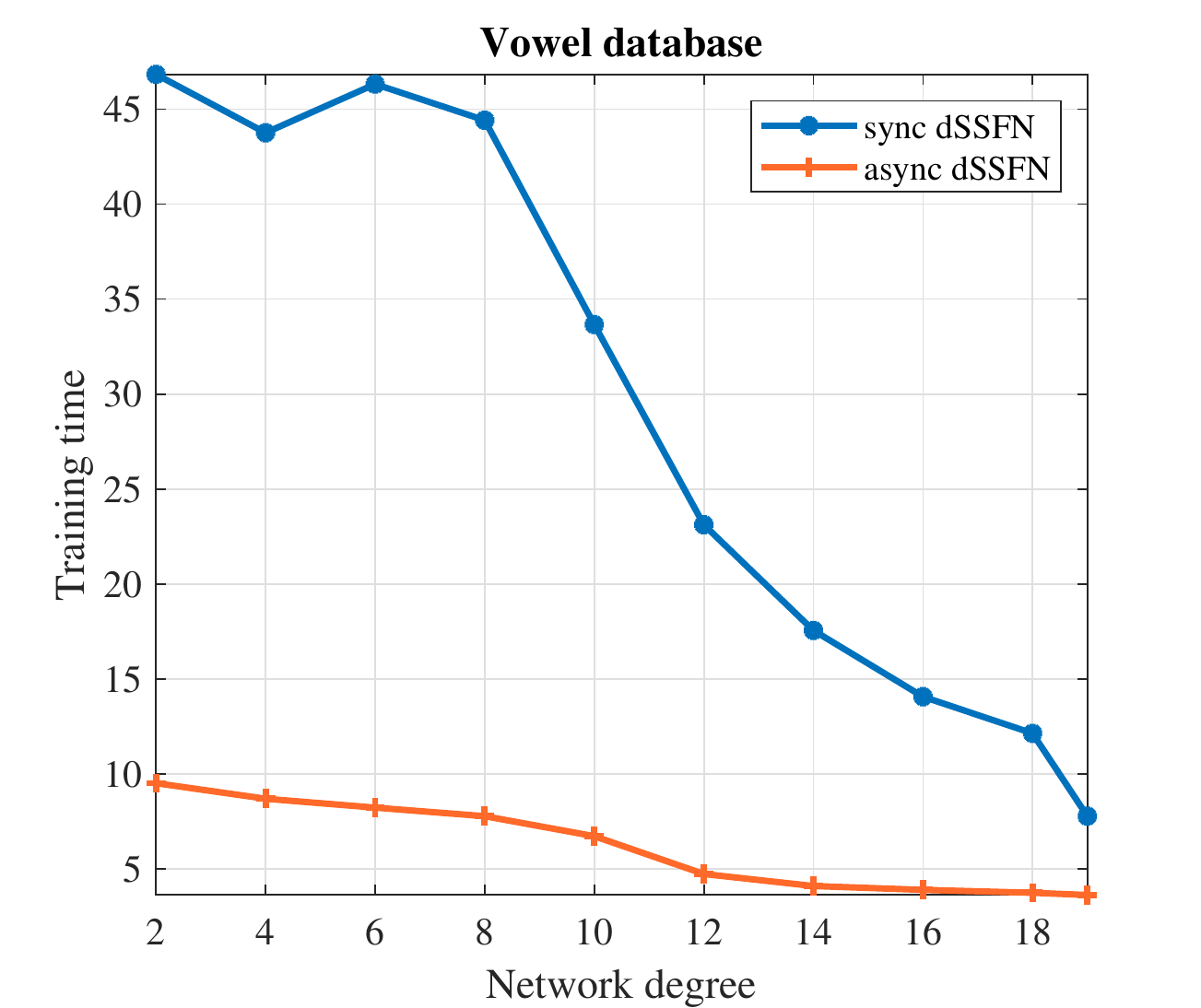}   
		\vspace{-5pt} 
		\caption{Training time versus network degree, $M=20$.}
		\vspace*{-5pt}
		\label{fig_II}
	\end{figure}

	\noindent\textbf{Effect of network degree.} We fix the number of nodes $M=20$, and test the performance of synchronous dSSFN and asynchronous dSSFN on different network degree $d$. The training time for different network degree is shown in Figure \ref{fig_II}. When the connections in the network increases, the traditional synchronous dSSFN speeds up because of the heavily reduced network consensus times in every ADMM iteration. While asynchronous dSSFN is not affected as much as synchronous dSSFN. This is consistent with the intuition that ARock based ADMM allows one side activation and communication (relaxed communication condition) so that the connectivity of network does not affect asynchronous dSSFN so much.    
	
\begin{figure}[t!]
		\centering
		        \vspace*{-5pt}
		\includegraphics[width=0.4\textwidth]{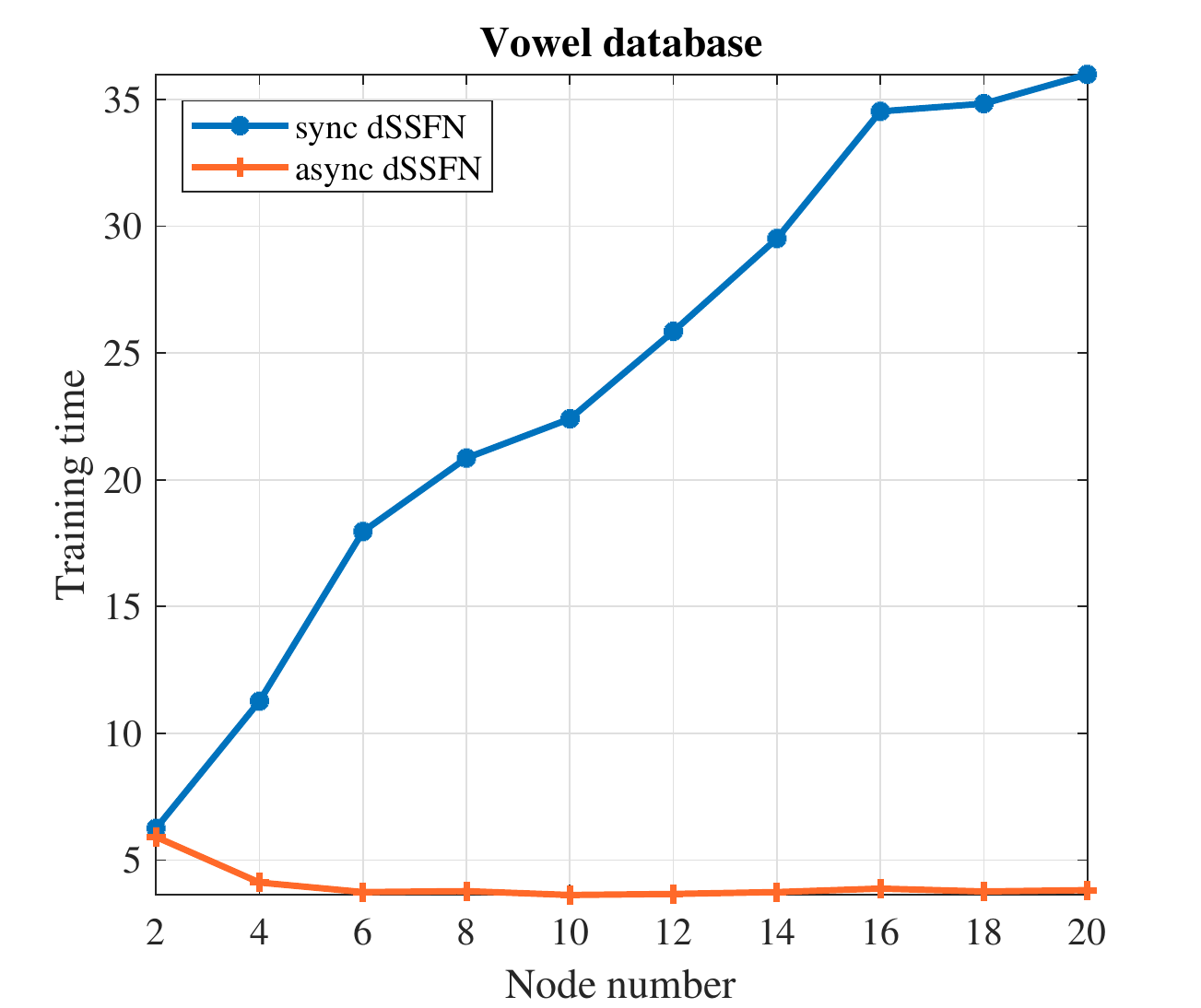}  
		\vspace{-5pt}  
		\caption{Training time versus node number, ring network.}
		\vspace*{-10pt}
		\label{fig_III}
	\end{figure}

	\noindent\textbf{Efficient computation and communication.} Figure \ref{fig_III} shows the training time versus number of processing nodes when the network degree is fixed to $d=2$ ($d=1$ for two-node network). This kind of network is also called a ring network. While the number of processing nodes increases, the training time of synchronous dSSFN increases monotonically due to the massive information exchange for network consensus in every ADMM iteration. But asynchronous dSSFN speeds up slightly when processing nodes increases. This implies that there is a trade-off between parallel computation benefit and communication overhead. Compared with synchronous dSSFN, asynchronous dSSFN enjoys a relaxed condition of communication by allowing one side communication and releasing nodes from waiting for up-to-date information from neighbors.
	
\section{Conclusion}

We have proposed an ARock based ADMM for solving the decentralized convex optimization in progressively learning the weight matrices of SSFN. We conclude that the proposed asynchronous dSSFN is efficient in computation and communication, which relaxes the condition of activation and communication. Therefore asynchronous dSSFN speeds up the learning process and remains competitive performance for sparse communication networks where the number of connected links is small.

	\bibliographystyle{IEEEbib}
	\bibliography{ICASSP2020}

\end{document}